\documentclass[prb,twocolumn,showpacs,groupedaddress]{revtex4}   
\pdfoutput=1
\usepackage{graphicx}  
\usepackage{dcolumn}   
\usepackage{bm}        
\usepackage{amssymb}   
\usepackage{times}

\begin{document}

\title{Glassy states in fermionic systems with strong disorder and interactions}
\author{David J. Schwab}
\author{Sudip Chakravarty}
\affiliation{Department of Physics and Astronomy, University of
California Los Angeles, Los Angeles, CA 90095-1547}
\date{\today}

\begin{abstract}
We study the competition between interactions and disorder in two dimensions.  Whereas a noninteracting system is always Anderson localized by disorder in two dimensions, a pure system can develop a Mott gap for sufficiently strong interactions.  Within a simple model, with short-ranged repulsive  interactions, we show that, even in the limit of strong interaction, the Mott gap is completely washed out by disorder for an infinite system for dimensions $D\le 2$, leading to a glassy state.  Moreover, the Mott insulator cannot maintain a broken symmetry in the presence of disorder.  We then show that the probability of a nonzero gap as a function of system size falls onto a universal curve, reflecting the glassy dynamics.  An analytic calculation is also presented in $1D$ that provides further insight into the nature of slow dynamics.
\end{abstract}

\maketitle 
\section{Introduction}
It has been a
dream in condensed matter physics to describe the quantum phase transition between 
localized and itinerant electrons, as it is a reflection of the basic concept of wave-partcle duality in a quantum many body system.  
Itinerancy mirrors the wave aspect, localization the particle aspect. In one-particle quantum mechanics, wave and particle descriptions are dual of each other, and there is no fundamental distinction between them. Coherent superposition of waves are packets that act like lumps of energy, or particles.  Yet, in a many particle system one believes that the metallic state, described by a non-normalizable wave function, is separated by a quantum phase transition from normalizable localized states, where particles are tied to spatial centers.

Band theory proposes a sharp distinction between metals and insulators.  Although a typical eigenstate
carries current, the totality of electrons in a filled band cannot. Paradoxically, in spite of the quantum
mechanically coherent and extended nature of each electronic eigenstate, the system is an
insulator.  The many
particle wave function of an insulator is a Slater determinant of Bloch functions of a filled
band. Alternately, the same  determinant can be rewritten as a
determinant of localized Wannier functions.  This is a manifestation of wave-particle duality.

An interaction driven insulator, or a Mott insulator,  can be an insulator even if the band is half-filled and can be due to a local repulsive, at most a few body, interactions.
While this can lead to a collective localized state,  this mechanism is vastly
different from the non-local statistical constraint enforced by the Pauli exclusion principle, as in a band insulator of non-interacting electrons.
Mott insulators are similar to classical insulators. Without quantum mechanics, at
zero temperature, a system of electrons will assume the configuration of the lowest potential energy due to  interactions, and 
because of the harmonic restoring force they will not conduct in response to an applied electric field. The lowest energy state is 
likely to be a broken symmetry state with crystalline order. 
A classical insulator is a localized state stabilized by  interactions.

There is another remarkable alternative, the Anderson insulator~\cite{andersonloc}. The non-interacting electronic
eigenstates may themselves localize due to a random potential, and if the Fermi energy is situated
within the localized states, the system is an insulator. Like a band insulator, quantum interference localizes a particle due to interference of time reversed paths, another manifestation of wave-particle duality.
A priori it is not clear when this physical situation realizes, as the role of  interaction becomes more and more important as the system approaches localization. Nonetheless, we would like to show that in certain circumstances the opposite may be true, that is, disorder dominates, however weak it may be. 

One of the mechanisms by which  an interacting system without disorder may become insulating is by opening a gap in the excitation spectra by breaking symmetries, such as spin and charge density waves that are particle-hole condensates. This mechanism provides {\em a} definition of Mott insulators, in the sense that a half-filled band could insulate.  Whether or not all Mott insulators {\it must} be accompanied by a broken symmetry has been the subject of some recent debate \cite{leeshankar,leeleinaas}.   Rather than addressing this issue, we shall {\it assume} that there is a broken symmetry in the Mott state, which is often the case, and in fact it is  the reason for its existence. We emphasize that broken symmetry is  a general concept for which  correlation effects  are {\em sine qua non}. Thus, the mechanism itself must not be identified with a Hartree-Fock approximation.  

The nature, and even the existence, of a $2D$ metal-insulator transition in low disorder Si-metal-oxide-semiconductor field-effect transistors has remained controversial since it was first reported \cite{kravchenko1}, despite considerable experimental and theoretical effort \cite{mitreview}.  The fundamental difficulty is in understanding the complex interplay between strong interactions and quenched disorder.  Noninteracting electrons (or even a Fermi liquid) are localized by any amount of disorder in two dimensions ($2D$) \cite{gangof4}, which implies that if a metallic phase is found in experiments, it must reflect a  non-Fermi liquid ~\cite{dobrosavljevic}.  

Due to their complexity, a principled analysis of systems involving both strong interactions and disorder is necessary to  understand what sorts of qualitative behaviors may result from these two basic ingredients. In this work, we will provide such an analysis, albeit in a simple model.  The paper is organized as follows. In Section II, we introduce the model and show, quite rigorously, that this model does {\it not} have a true metal-insulator transition in the presence of disorder.  The analysis is performed in the strong interaction limit where the Mott gap is the largest and should be the most resistant to the onslaught of disorder.  In Section III, we study, through numerically constructed ground states, the probability for a finite system to have a non-zero gap, and show that this quantity falls onto a universal curve.  From this result, we then show that the system possesses glassy quantum dynamics.  In Section IV, we then specialize to the case of one dimension and study analytically the disorder averaged ground state density as a function of chemical potential.  We compare this to the density achieved from a rapid temperature quench to illustrate anomalously slow dynamics even in $1D$.  Finally, we conclude in Section V with a brief discussion of the implications of our results.

\section{The Model}
Consider the simplest $2D$ fermion model  that has a broken symmetry in the insulating state: spinless fermions on a square lattice with the Hamiltonian

\begin{equation} \label{eq:hamiltonian}
H_p=-\frac{1}{2} \sum_{<i,j>}  \left[c^\dagger_ic_j+\mathrm{h.c.} \right]+\Delta \sum_{<i,j>}(n_i-\frac{1}{2}) (n_j-\frac{1}{2})
\end{equation}
The sums are over  nearest neighbors of a  bipartite lattice with sublattices $A$ and $B$.  The symmetry under the operation $c_i\rightarrow  - c_i^\dagger$ in $A$, while $c_i\rightarrow  c_i^\dagger$ in $B$,  ensures half-filling, with $n_{i}=c^{\dagger}_{i}c_{i}$ being the density operator at site $i$.  
This model is well studied~\cite{Luther:1975,Baxter:1982} in $1D$ where the system is a Mott localized insulator in our sense, a charge density wave state, with a gap, $g$, for  $\Delta>1$; $g\to \Delta$, for $\Delta \gg 1$.  For $\Delta<1$, the system is metallic, therefore $\Delta=1$ is the location of a metal-insulator transition. The same transition must obviously be present in $2D$ on general grounds, as the fluctuations are weaker in higher dimensions.

\begin{figure}[htb]
\begin{center}
\includegraphics[scale=0.2]{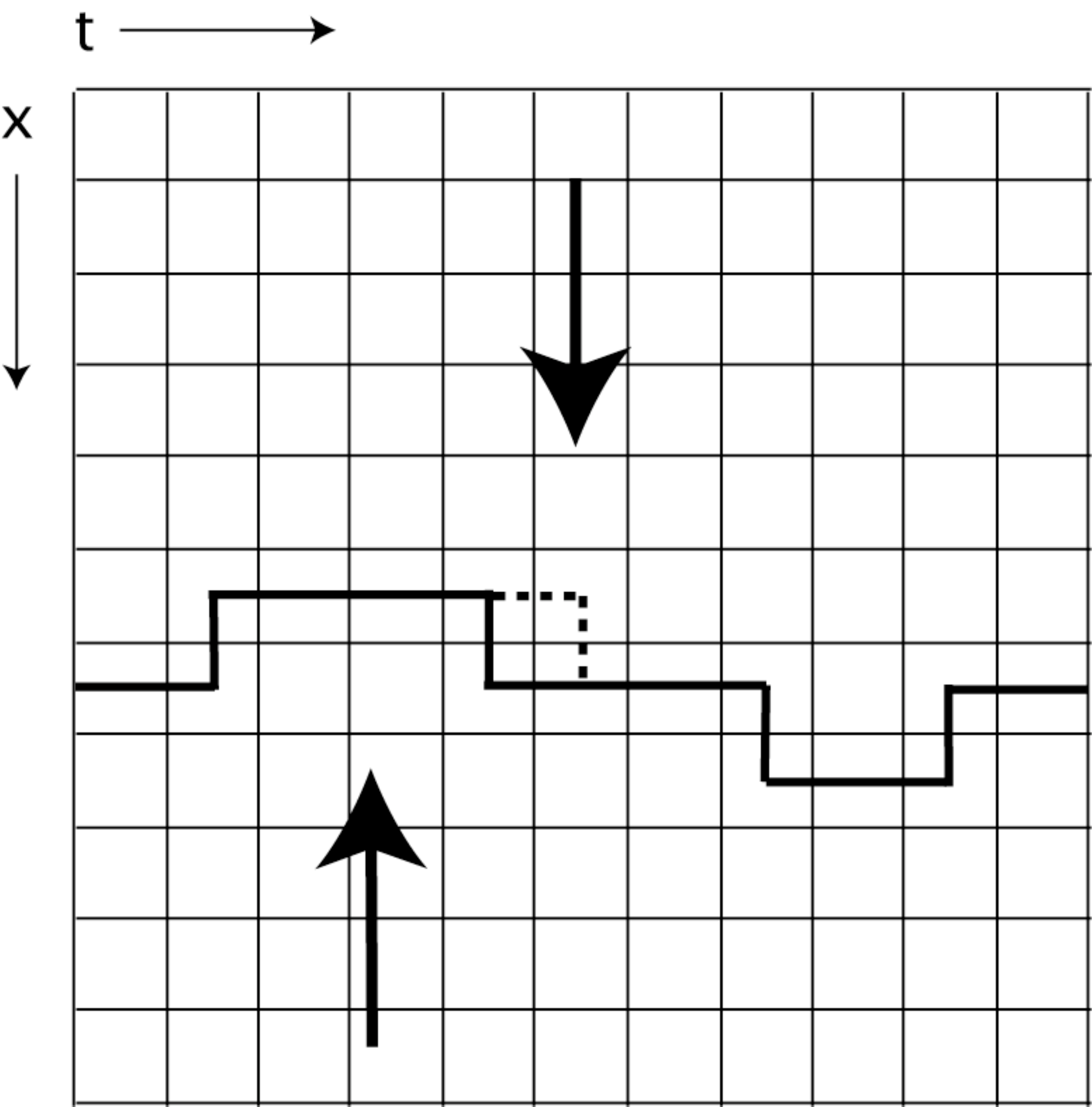} 
\end{center}
\caption{Example of a domain wall separating a region of down spins (above) from up spins (below).  The dotted line is an example of a (gapless) corner excitation in which no extra bonds are broken.}
\label{fig:corner}
\end{figure}

Addition of a random potential gives $H=H_p+H_r$ with
$
H_r=- \sum_{i} V_i n_i .
$
where $V_i$ are independent Gaussian random variables with zero mean and variance $\sigma^2$.  
 Consider the case where we drop the hopping terms and we are left simply with a classical model since the $\Delta$ term clearly commutes with $H_r$.  {\em We do this not as an approximation, but to prove the point that even in the limit that the Mott insulator has the best chance of surviving, any amount of randomness destroys the Mott gap.}   We are strongly motivated by an argument offered for $1D$~\cite{shankar}. As $\Delta$ is lowered, the kinetic energy will cause the walls to fluctuate, but this won't change the fact that the gap has been destroyed by randomness, nor will it restore the symmetry.    {\em  As long as the disorder remains finite, it is difficult to believe that the system will ever reach a true metallic state. }

Using $n_i=S_z(i)+1/2$, we see that we have, in fact, an Ising model.  Redefining $S_z(i)\rightarrow -S_z(i)$ at every other site, and remembering that the $V_i$ are symmetrically distributed with zero mean, gives simply the classical ferromagnetic random field-Ising model (RFIM):

\begin{eqnarray} \label{eq:rfising}
H= -\Delta \sum_{<i,j>} S_z(i)S_z(j)-\sum_{i} V_i S_z(i).
\end{eqnarray}

Although $2D$ is the marginal dimension for the Imry-Ma \cite{imryma} argument, it has been rigorously shown  by Aizenmann and Wehr~\cite{aizenmannwehr} that there is no long-range order for arbitrarily weak randomness.  Moreover, since the ground state will generically be a disordered spin configuration dependent upon the particular realization of the random potential, the excitations of the system result from moving domain walls between the up and down spin regions.  As a result of the disorder, the domain walls will in general be rough, and the elementary excitations then consist of moving corners because no extra bonds are broken (see Fig.~\ref{fig:corner}).  The energy cost of moving corners depends only on the random field configuration and {\it not} on $\Delta$.  Thus, for an infinite system, the excitation spectrum will be essentially {\it gapless} for a continuous distribution of $V_i$, which we have assumed.  Note that the corner excitations are not necessarily the first excited states, but they are sufficient to prove the existence of gapless excitations. 

\section{Two-Dimensional Ground States and Glassy Dynamics}
Having argued against the existence of a metal-insulator transition in the presence of disorder, we turn to an analysis of the ground state in $2D$ and provide evidence for glassy quantum dynamics.  As already mentioned, $D=2$ is special for the RFIM.  In $1D$, the Imry-Ma argument easily gives that ordered domains have size $L_c\sim\left(\frac{\Delta}{\sigma}\right)^2$.  But since $D=2$ is the marginal dimension, no information can be gleaned from the simple Imry-Ma argument.  Rather, it is necessary to study the energy gained upon allowing the domain walls to roughen.  Such a calculation was performed by Binder \cite{binder} with the result, \cite{binder,natterman}, that $L_c\sim \exp\left[{A \left(\frac{\Delta}{\sigma}\right)^2}\right]$, where A is a constant.  Thus, domains in two dimensions are {\it exponentially} larger than those in $1D$.  It is important to note that this equation relates the ${\it typical}$ size of domains to the ratio of exchange to random field energies. 

This leads us to ask the following question: Given a finite system of size $L\times L$, what is the probability it possesses an energy gap?  For small $L\ll L_c$, a non-zero gap would be fairly likely, but as $L$ increases towards the crossover length, $L_c$, given above, the probability of finding a gap should decrease because the system may now contain multiple ordered domains separated by rough walls.
  
\subsection{Numerically Computed Ground States}
\begin{figure}[htb]
\begin{center}
\includegraphics[width=5cm]{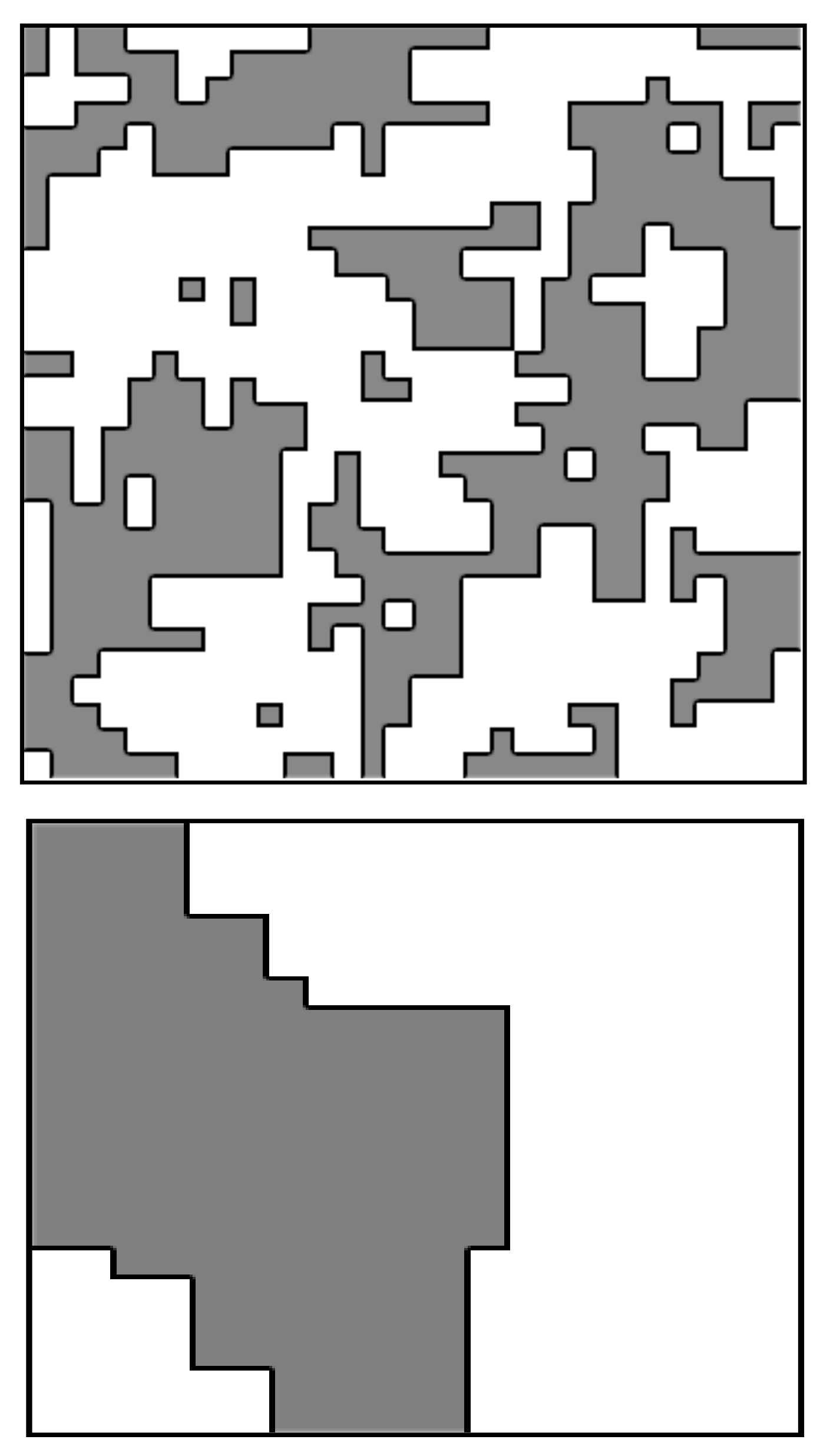} 
\end{center}
\caption{Typical ground state configurations for different disorder strengths and system sizes.  Top: $L=30\times30$ and $\Delta/\sigma=3$.  Bottom: $L=20\times20$ and $\Delta/\sigma=.95$.}
\label{fig:domain}
\end{figure}

We answer this question numerically by computing exact ground state configurations of the RFIM for different system sizes and disorder strengths.   The general method \cite{ogeilski,hartmann} for finding ground states of random field systems (or even random bond systems without frustration), is based on a mapping to an equivalent minimum cut network flow problem.  In a network, nodes (i.e. lattice sites) are connected by directed links with finite capacity, signifying the maximum possible flow between neighboring nodes.  Two additional sites (dubbed the source and sink) are augmented to the lattice, and each site of the RFIM is connected by a directed link to one of the two external sites, depending on the sign and strength of its random field.  The flow capacity between neighboring lattice sites is determined by the exchange energy.  Dividing the network in two, with source and sink on opposite sides of the division, defines a cut.  With this construction, the minimum of the capacity across all possible cuts provides the ground state energy of the RFIM, while the minimum cut itself determines the spin configuration.  Using the equivalence between the minimum cut capacity and the maximum flow through the network, known as the max flow-min cut theorem~\cite{Ford:1956}, allows one to simply calculate maximum flows.  To do this, we employed the efficient push-relabel code \cite{goldbergcode} which enables us to get good statistics for moderate system sizes.  To illustrate typical ground state domain structures for various disorder strengths, we computed minimum cuts explicitly through the Edmonds-Karp algorithm \cite{EdmondsKarp}, which are displayed in Fig.~\ref{fig:domain}.
        
\begin{figure}[ht]
\begin{center}
\includegraphics[width=\linewidth]{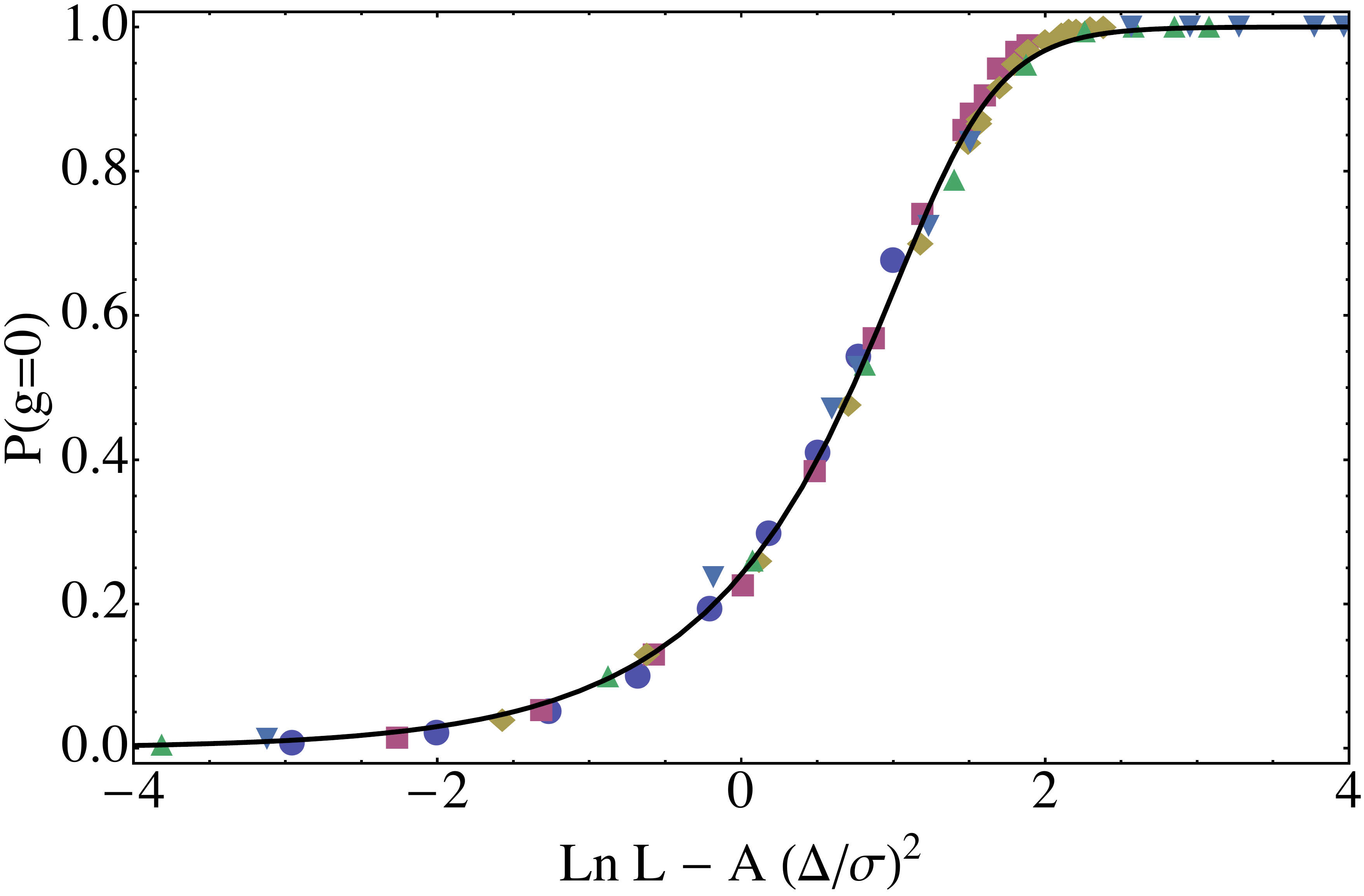} 
\end{center}
\caption{(Color online) Probability to find an energetically favorable domain wall in a finite system of linear size $L$.  The values of $\Delta/\sigma$ range from 0.2 to 1, while the values of $L$ are 20, 40, 80, 160, and 320. Symbols represent different fixed system sizes with varying disorder strengths.  For each choice of $\Delta/\sigma$ and $L$, we averaged over 3,000 realizations of disorder.  We find that $A=1.8$ provides the best data collapse.  The black line is the fit to an asymmetric sigmoid (see text).}
\label{fig:scaling}
\end{figure}

For each
chosen value of $\sigma/\Delta$ and $\ln L$, we calculate the fraction of
realizations of the $V_i$ that contain a domain wall in their ground
state. To detect a  domain wall, the exact ground state energy computed via the network flow model is compared with the minimum energy of the two ferromagnetic states (all up or all down). If the ground state  energy is lower, there must be a domain wall.  If
not, and the ground state energy equals the lower energy ferromagnetic state, then
there must not be a domain wall. This procedure avoids  having to examine the spin
configuration explicitly.   In other words, we will assume that the presence of a domain wall implies gapless excitations even for a finite system.  On the other hand, if the ground state is purely ferromagnetic, the excitation energy will be non-zero and of order $\Delta$.  Figure~\ref{fig:scaling}  shows our results.  
As expected, the probability of domain wall formation grows upon increasing both $\ln L$ and $\sigma/\Delta$.  The surprising feature is that when we define the $x$-axis to be $x=\ln L - A \left(\frac{\Delta}{\sigma}\right)^2$, all points collapse onto a single universal curve.  We find $A=1.8$, in good agreement with a previous study which reported $A=2.1\pm.2$  \cite{seppala} at the special value $P(g=0)=1/2$.  The collapse onto a single curve enables us to gain information about the slow transition to a disordered ground state even for values of  $\sigma/\Delta$ for which the relevant system sizes are far too large to be studied numerically.  Most notably, not only does the ${\it typical}$ size for observing a disordered ground state (i.e. the $L$ for which $P(g=0)=0.5$) scale as predicted in \cite{binder}, the ${\it entire}\,{\it distribution}$ scales in precisely the same way.  This is surprising because it might have been expected that systems with particularly weak randomness cross over to disordered ground states more `slowly', i.e. over a much broader range of $\ln L$, in addition to reaching $P(g=0)=0.5$ at a larger length scale.  Note also that the gap probability is also {\it not} symmetric about $P(g=0)=0.5$.

The crossover from a generically ordered $\left(P(g=0)\sim0\right)$ to disordered $\left(P(g=0)\sim1\right)$ ground state occurs quite slowly, over nearly two decades.  The large range of $L$ for which samples are neither generically ordered nor disordered indicates the possibility of large sample to sample fluctuations.  Some systems may have remnants of a Mott gap, while others are localized by disorder.  In addition, the large $L$ regime will likely possess many low-lying energy states, each requiring the reorganization of large numbers electrons, implying significant metastability and glassy dynamics, akin to experiments on $2D$ low-mobility Si inversion layers \cite{Bogdanovich,Jaroszynnski:2002,jaroszynski:2007,jaroszynski:2007b}, although the interactions considered here are short-ranged.  Such glassy behavior has also been found theoretically in a similar model of spinless fermions on a Bethe lattice \cite{glassy}. It is quite remarkable that contrary to expectations the system behaves more like an ``Anderson insulator'' despite strong interactions.

\subsection{Glassy Dynamics}
The signature of glassy quantum dynamics lies in the long time-scale for relaxation to the ground state.  We can, in fact, learn something about the dynamics by interpreting Figure 3.  The distribution $P(g=0)$ can be regarded as the cumulative distribution function (cdf) to find an ordered domain of characteristic size $L$ or smaller, since this distribution function reflects the existence of a domain wall up to the scale $L$.  The  curve fits an asymmetric sigmoid $f(x)$ of the form
\begin{equation}
f(x) = \frac{1}{\left(1+e^{(x_{0}-x)/\xi}\right)^{\theta}}, 
\end{equation}
where $x=\ln(Le^{-A(\Delta/\sigma)^{2}})$. The best fit to the data shown in Fig.~\ref{fig:scaling}  is $\theta=0.31$, $x_{0}=1.37$, and $\xi=0.29$.   Taking a derivative to find the $L$ distribution, $P(L)$, results in $P(L)\sim L^{-1/\xi}$ for large $L$.  
If we define the imaginary part of the frequency dependent local susceptibility corresponding to the density $n_{i}(t)$ (Heisenberg operator) to be $\chi''(\omega)$, then 
\begin{equation}
\chi''(\omega)\sim \int dL \; P(L) \; \delta(\omega - \omega_{0}e^{-c L^{\alpha}}).
\end{equation}
The $\delta$-function signifies that at a frequency $\omega$ the quantum tunneling rate corresponding to  that frequency is sampled by a cluster of size $L$, where the exponent $\alpha$ requires microscopic calculation and is left undetermined in the present phenomenological analysis. The quantity $\omega_{0}$ is the attempt frequency in the many dimensional WKB theory. Thus, it is easy to show that as $\omega \to 0$,
\begin{equation}
\chi''(\omega)\sim \frac{1}{\omega}\frac{1}{(\ln \frac{\omega_{0}}{\omega})^{\psi}}.
\end{equation}
The exponent $\psi= 1+(1/\xi-1)/\alpha> 1$, as long as $\alpha\ne 0$; $\alpha=0$ is highly unlikely because that would imply that the action corresponding to the tunneling rate is independent of the size of the cluster, $L$. It should be interesting to check experimentally that the noise power spectrum does follow this $1/\omega$-law with a logarithmic correction, signifying glassy dynamics.

\section{Analytic Results in One Dimension}

It is instructive to consider the same model of spinless fermions in one dimension~\cite{shankar} where it is possible to compute disorder averages of thermodynamic quantities analytically.  Recall that in $1D$, the Imry-Ma argument gives that the ordered domains have characteristic size $L_c\sim\left(\frac{\Delta}{\sigma}\right)^2$, so the ground state is "disordered" and heterogeneous.  To illustrate the origin of glassy dynamics present even in $1D$, we will calculate both the disorder averaged ground state density $\rho_G$ as well a quenched density $\rho_Q$ that would be obtained from an infinitely fast temperature quench.  This quenched state is obtained through the sequential filling of the lowest available energy levels up to the chemical potential $\mu$.  We will be interested in how the density profiles vary with $\mu$, so we must relax the condition of half-filling.  In addition, we will work in the limit of large nearest-neighbor repulsion, $\Delta \rightarrow \infty$, so that the particles must be separated by at least one empty lattice site.  In other words, the fermions may be regarded as hard-core dimers (see  Fig. \ref{fig:qvsg1}).  We emphasize that these dimers should not be confused with valence bonds connecting two neighboring sites, for which the word "dimer" is frequently also used.  The hard-core dimer constraint, along with the random on-site energies, induces a geometrical frustration between competing particle configurations.  As a result, the ground state density profile is a complex structure that incorporates the preference for low energies while still respecting the hard-core constraint.

\subsection{\label{heuristics} Heuristic Analysis of Dimer Frustration}
Before calculating $\rho_G$ and $\rho_Q$, we give a simple argument to show that these two quantities differ by a finite amount for {\it all} $\rho_G>0$.  In what follows, the on-site energies will be drawn from a uniform (i.e. rectangular) distribution, denoted by $R(\epsilon)$, between zero and one.  The uniform distribution will simplify the analytics and preserve the consistency of the dimer (i.e. large $\Delta$) limit.  When $\mu$=0, all sites have positive energy, so the ground state is an empty lattice.  When $\mu=1$, all sites are attractive, but the ground state has a complicated structure due to the competing effects of the random on-site energies and the hard-core constraint.  Therefore, we will focus on $\mu$ between zero and one.  In the following argument, we will absorb $\mu$ into the on-site energies which will instead be uniformly distributed between $-\mu$ and $1-\mu$.  

\begin{figure}[t]
\begin{center}
\includegraphics[scale=0.27]{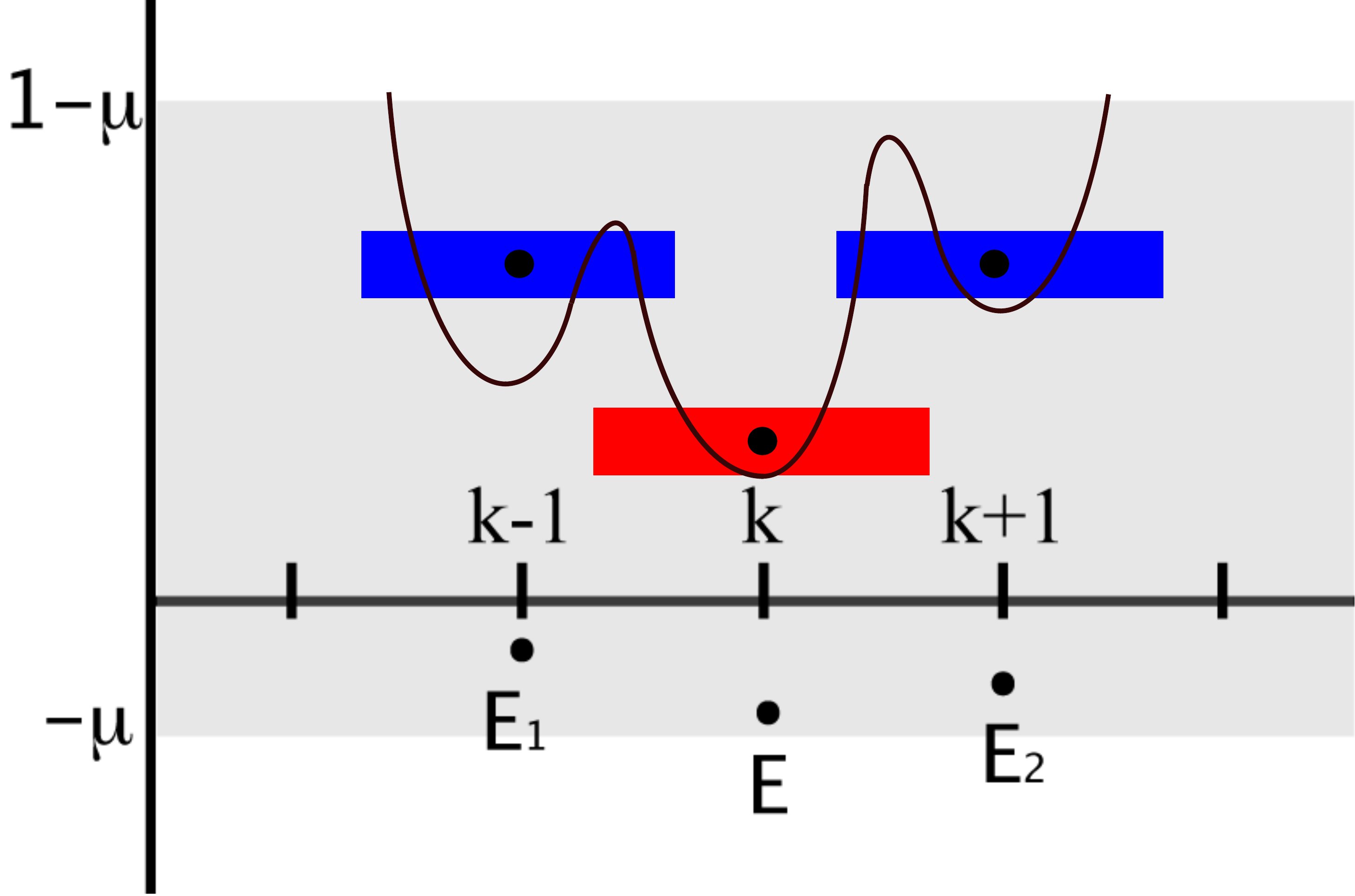} 
\end{center}
\caption{(Color online) Schematic of a scenario where the quenched configuration occupies site $k$, hence blocking sites $k-1$ and $k+1$, but the ground state foregoes occupancy of site $k$ in favor of the flanking sites if $E_1+E_2<E$.  The upper (blue) configuration is the ground state and the lower (red) is the quenched.}
\label{fig:qvsg1}
\end{figure}

Consider a finite lattice of $L$ sites.  Let the energy at site $k$ be $E<0$ and the energies at sites $k-1$ and $k+1$ be $E_1$ and $E_2$, respectively.  We calculate the probability, $P_{\mbox{switch}}$, that in the ground state, placement of a particle at $k$ is forfeited in favor of the occupation of sites $k-1$ and $k+1$, despite site $k$ having the lowest energy.  The scenario is depicted in Figure \ref{fig:qvsg1}.  This is the simplest way geometric frustration may cause the ground and quenched states to differ because $\rho_G(k)=0$ but $\rho_Q(k)=1$.  Averaging over the value of $E$, 

\begin{eqnarray}
P_{\mbox{switch}}=-\frac{\int_{-\mu}^0 P(E_1+E_2<E)\mathrm{d}E}{\int_{-\mu}^0 \mathrm{d}E} 
\label{switch}
\end{eqnarray}
Since the sites $k-1$ and $k+1$ must also be attractive, $P(E_1+E_2<E)=\int_E^0 dE_1\int_E^0 dE_2 \theta\left(E-(E_1+E_2)\right)$ which equals $E^2/2$.  Plugging this into (\ref{switch}) then gives that $P_{\mathrm{switch}}=\mu^2/6$.  Multiplying $P_{\mathrm{switch}}$ by the average number of sites with $E<0$, i.e. $\mu L$, gives the average number of these switches in a finite system of size $L$.  Setting this equal to unity gives the chemical potential at which we expect the first switch: $\mu_c \sim L^{-1/3}$.  Clearly, as $L\rightarrow \infty$, $\mu_c \rightarrow 0$.  Also, since each such switch increases the ground state density relative to the quenched, we also have that $\delta=\rho_G-\rho_Q\sim \mu^3/6$ for small $\mu$, so $\delta=0$ only at $\mu=0$.  The above argument neglects the contributions of sites $k-2$ and $k+2$ etc. but these effects are higher order in $\mu$ and hence can be neglected for $\mu\ll1$.  We will see this behavior of $\delta$ reproduced precisely from the exact result.

\subsection{\label{groundstate} Exact Solution for Ground State Density}
We now derive the disorder-averaged ground state density $\rho_G$ for a $1D$ lattice of dimers at chemical potential $\mu$ and with on-site energies uniformly distributed between zero and one.  The analysis follows that of Fonk and Hilhorst \cite{fonk} who considered the problem with a different energy distribution not easily generalizable to a non-zero chemical potential.  Define $E_k^1 (E_k^0)$ to be the ground state energy of the first $k$ sites subject to the constraint that a particle is present (absent) at site $k$.  These quantities can easily be seen to obey the recursion relations,

\begin{eqnarray}
E_k^1=\epsilon_k+E_{k-1}^0\\
E_k^0=\min \left(E_{k-1}^0,E_{k-1}^1\right)
\end{eqnarray}
where $\epsilon_k$ is the random energy at site $k$.  Defining difference variables $\xi_k=E_k^1-E_k^0$ and subtracting the above equations gives a simple recursion relation for $\xi_k$,

\begin{eqnarray}
\xi_k=\epsilon_k+\min\left(0,-\xi_{k-1}\right)
\end{eqnarray}

By averaging over the on-site energy distribution $R(\epsilon)$, the recursion relation is readily transformed into an integral recursion relation for $P(\xi)$, the distribution function of $\xi$,

\begin{eqnarray}
P_k(\xi)=R(\xi)\int_{-\infty}^0\mathrm{d}\xi' P_{k-1}(\xi')\\
+\int_{0}^{\infty} \mathrm{d}\xi' R(\xi+\xi') P_{k-1}(\xi')\nonumber
\end{eqnarray}

The fixed point distribution of $P(\xi)$ will contain the required information for bulk quantities, so we can drop the subscripts on $P(\xi)$.  With $R(\epsilon)=\theta(\epsilon+\mu)\theta(1-\mu-\epsilon)$, we see that $P(\xi)=0$ for $\xi>1-\mu$ and hence also for $\xi<-1$.  Then there are three distinct regions to consider:

\leftline{}
{\bf Region 1: }$ -1<\xi<-\mu $

\begin{eqnarray}
P(\xi)=\int_{-\xi-\mu}^{1-\mu}P(\xi')\mathrm{d}\xi' 
\end{eqnarray}

{\bf Region 2: }$-\mu<\xi<0$

\begin{eqnarray}
P(\xi)=\int_{-1}^{0}P(\xi')\mathrm{d}\xi' +\int_{0}^{1-\mu}P(\xi')\mathrm{d}\xi' =1
\end{eqnarray}

{\bf Region 3: }$0<\xi<1-\mu$

\begin{eqnarray} \label{eq:integral}
P(\xi)=\int_{-1}^{0}P(\xi')\mathrm{d}\xi' +\int_{0}^{1-\mu-\xi}P(\xi')\mathrm{d}\xi' 
\end{eqnarray}

If we know the solution in region 3, we can integrate to find the solution in region 1.  Region 2 has a flat value of 1 (since $P(\xi)$ is a normalized probability distribution).  We convert (\ref{eq:integral}) to the differential equation

\begin{eqnarray}
\frac{dP(\xi)}{d\xi}=-P(1-\mu-\xi)
\end{eqnarray}
which can be reduced to two coupled ODEs by the replacement $Q(\xi)=P(1-\mu-\xi)$.  The resulting solution for $P(\xi)$ in region 3 is

\begin{eqnarray} \label{eq:solution1}
P(\xi)=\cos \xi+\frac{\sin \left( \frac{1-\mu}{2}\right)-\cos \left( \frac{1-\mu}{2}\right)}{\sin\left( \frac{1-\mu}{2}\right)+\cos\left(\frac{1-\mu}{2}\right)}\sin\xi
\end{eqnarray}

There is an undetermined multiplicative constant fixed by requiring that $P(\xi)$ integrates to 1.  From the form of the equation in region 3, we see that this is equivalent to requiring that $P(1-\mu)+\int_0^{1-\mu}P(\xi')\mathrm{d}\xi'=1$.  The necessary constant turns out to be unity.  One can then directly integrate to find the solution in region 1:

\begin{eqnarray} \label{eq:solution2}
P(\xi)=\frac{2 \sin \left( \frac{1+\xi}{2}\right)\left[\sin \left( \frac{\xi+\mu}{2}\right)+\cos \left( \frac{\xi+\mu}{2}\right)\right]}{\sin\left( \frac{1-\mu}{2}\right)+\cos\left(\frac{1-\mu}{2}\right)}
\end{eqnarray}

Eqs. (\ref{eq:solution1}) and (\ref{eq:solution2}), along with $P(\xi)=1$ in region 2, comprise the required solution of the integral equation.

We now use the derived form of $P(\xi)$ to solve for the disorder-averaged ground state density.  This can be found, again following \cite{fonk}, by defining $E^0 (E^1)$ to be the minimum energy of the {\it entire} system (not just the left half), subject to the constraint that a particle is absent (present) at some site $k$ deep in the bulk.  The average density will be given by $1-P(E^0<E^1)$.  A similar recursive calculation, this time including sites to the left and right, leads to

\begin{eqnarray}
P(E^0<E^1)=\int_{-1}^{1-\mu}\mathrm{d}\xi_1P(\xi_1)\int_{-1}^{1-\mu}\mathrm{d}\xi_2P(\xi_2) \times\\
\int_{-\mu}^{1-\mu}\mathrm{d}\epsilon \theta\big(-\epsilon-\min[0,-\xi_1]-\min[0,-\xi_2]\big)\nonumber
\end{eqnarray}
The theta function can be split up into four cases

\begin{eqnarray}
\lefteqn{\int_{-\mu}^{1-\mu}\mathrm{d}\epsilon \theta\big(-\epsilon-\min[0,-\xi_1]-\min[0,-\xi_2]\big)=} \\
& &\big[\theta(\xi_1)\theta(-\xi_2)+\theta(-\xi_1)\theta(\xi_2)\big](\xi_1+\mu)+\nonumber \\
& &\theta(\xi_1)\theta(\xi_2)\min(1,\xi_1+\xi_2+\mu)+\theta(-\xi_1)\theta(-\xi_2)\mu\nonumber
\end{eqnarray}
and each term integrated with the form of $P(\xi)$ derived above.  After some lengthy but straightforward algebra, and using $\rho_G=1-P(E^0<E^1)$, we find the remarkably simple result 

\begin{eqnarray}\label{eq:ground}
\rho_G=\frac{1}{1+\csc\mu}
\end{eqnarray}
It is important to remember that this result is valid in the regime $0\leq\mu\leq1$.  Other quantities such as the average energy per site can be calculated from $P(\xi)$, should they be of interest.

\subsection{\label{quenchedstate} Quenched State Density}
We now turn to the calculation of the quenched state density.  To do this, we will use the formalism of random sequential adsorption (RSA). \cite{rsareview}  In particular, we will use the dynamic formulation of RSA with a random distribution of binary adsorption rates \cite{stacchiola} in the determination of the quenched density.  The reason is that, since the random energies on distinct sites are uncorrelated, the process of sequentially filling the deepest energy minima is identical to an RSA process \cite{fonk}.  However, with $\mu<1$, some sites are repulsive and hence have an "on rate" of zero.  Therefore, we need to consider an RSA process with two adsorption rates, $\alpha$ and $\beta$, take the limit $\beta\rightarrow0$ (while $\alpha$ remains arbitrary), and look for the $t\rightarrow\infty$ density.

\begin{figure}[t]
\begin{center}
\includegraphics[width=\linewidth]{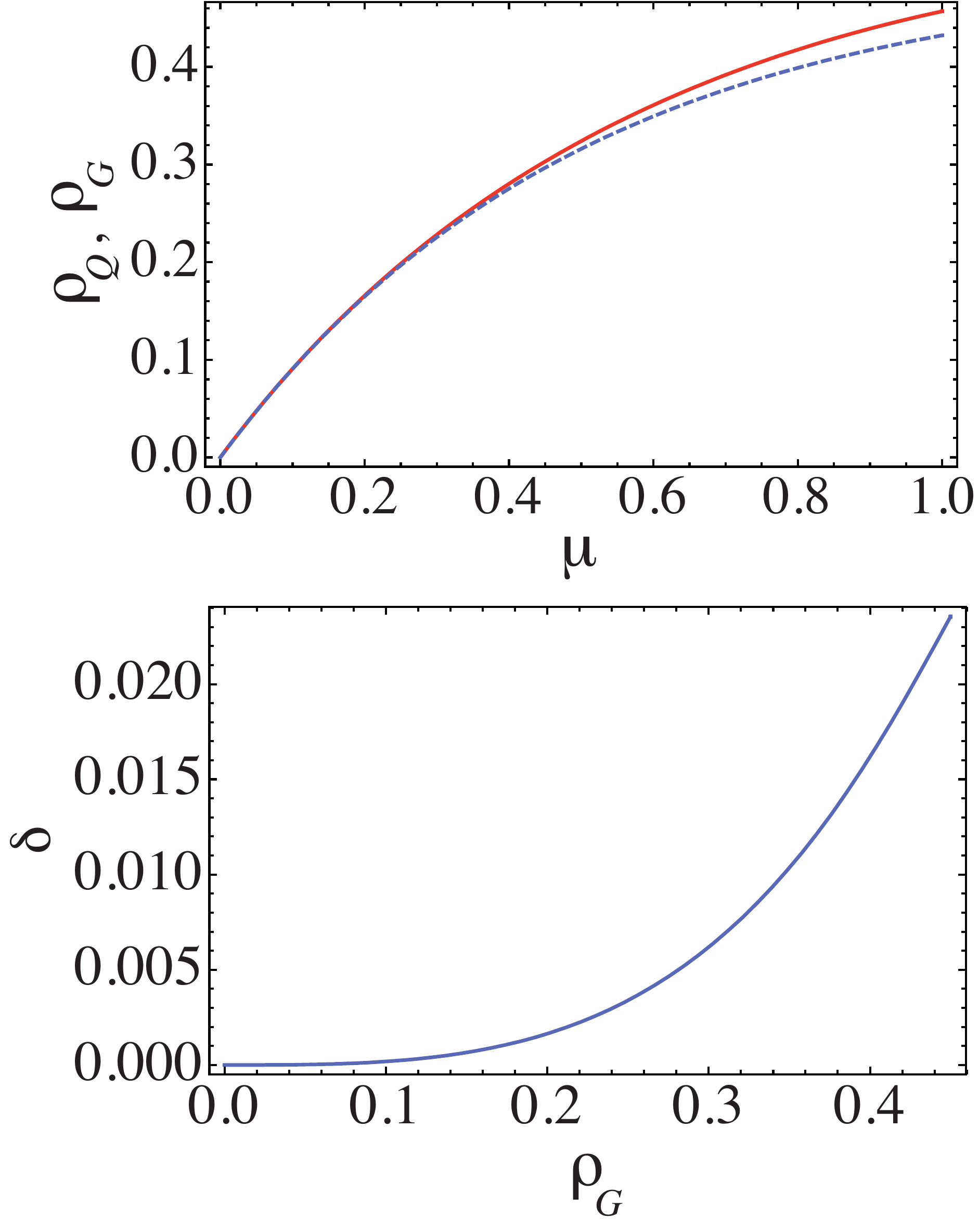} 
\end{center}
\caption{(Color online) (A) Plot of the analytic forms of the disorder averaged ground (red solid line) and quenched (blue dashed line) state densities.  Both initially rise linearly from zero, but the quenched density peels off for higher $\mu$.  Note that $\rho_G$ is bounded below by $\rho_Q$ and the two functions only intersect at zero.  (B) Plot of $\delta=\rho_G-\rho_Q$ vs. $\rho_G$.}
\label{fig:qvsg2}
\end{figure}

The adsorption rate of site $n$ will be denoted by $\alpha_n$ and the probability that site $n$ is occupied by a particle at time $t$ is $\rho_n(t)$.  This probability should be thought of as an average over different realizations of the adsorption process for a fixed choice of the $\alpha_n$.  Using established formalism\cite{stacchiola}, $\rho_n$ varies in time as

\begin{eqnarray}
\frac{d\rho_n(t)}{dt}=\alpha_n\exp(-\alpha_n t)Q^-_{n+1}Q^+_{n-1}
\end{eqnarray}
where the $Q$'s are time-dependent and obey

\begin{eqnarray}
\frac{dQ^-_n}{dt}=-\alpha_n\exp(-\alpha_n t)Q^-_{n+1}
\end{eqnarray}
with a similar equation for $Q^+_n$ except with the replacement $n+1\rightarrow n-1$.  Since in our calculation of the ground state density, we have chosen the on-site energies to be uniformly distributed between zero and one, the chemical potential $\mu$ gives the fraction of attractive sites that therefore have adsorption rate $\alpha\neq0$.  Thus we choose the $\alpha_n$'s to be random variables equal to $\alpha$ with probability $\mu$ and equal to zero with probability $1-\mu$.  The rest of the sites are repulsive and hence have an adsorption rate of zero.  The sequential filling of lowest energy minima will then be mimicked by the dynamic RSA process.  

To find the average density of the quenched configuration, $\rho_Q$, we need to average over the $\alpha_n$'s and take the $t\rightarrow \infty$ limit.  Since $Q_{n+1}^-$ only depends on sites $m\geq n+1$ (and $Q_{n-1}^+$ only on $m\leq n-1$), the average over $\alpha_n$, denoted by $\left<.\right>$, simply factorizes \cite{stacchiola}:

\begin{eqnarray}\label{eq:qdynamics}
\frac{d\left<\rho_Q(t)\right>}{dt}=\alpha \exp(-\alpha t)\left<Q\right>^2\\
\frac{d\left<Q\right>}{dt}=-\mu \alpha \exp(-\alpha t)\left<Q\right>
\end{eqnarray}
where we have used $\left<\alpha_n\exp(-\alpha_n t)\right>=\mu \alpha \exp(-\alpha t)$ and the fact that the $Q$'s become independent of position after averaging, due to translational invariance.  Clearly, we then have $\left<Q\right>=\exp\left[\mu (e^{-\alpha t}-1)\right]$ and upon integration of (\ref{eq:qdynamics}) for $t\rightarrow \infty$, we find that the quenched density is given by

\begin{eqnarray}\label{eq:quenched}
\rho_Q=\frac{1}{2}(1-e^{-2 \mu})
\end{eqnarray}
$\rho_Q$ rises linearly from zero at $\mu=0$ and saturates at $\mu=1$ to the "jamming" density of dimers $\simeq .432$.

\subsection{Discussion}
The two densities (\ref{eq:ground}) and (\ref{eq:quenched}) are plotted in Figure \ref{fig:qvsg2}, and their difference $\delta$ is plotted below.  $\rho_Q$ provides a lower limit for $\rho_G$ and both rise linearly from zero for small values of $\mu$.  For larger $\mu$, $\rho_Q$ peels off due to the jamming caused by irreversible adsorption.  When we expand $\delta=\rho_G-\rho_Q$ for small $\mu$, the first non-zero term is $\mu^3/6$, identical to what was predicted earlier in our heuristic argument.  We also plot $\delta$ vs. $\rho_G$ in Figure \ref{fig:qvsg2} to show that $\delta$ remains quite small until the lattice reaches quarter filling ($\rho_G=.25$).  

The picture that emerges from our exact solution leads to a few general conclusions.  First, the quenched and ground states of hard core particles in disordered landscapes {\it always} differ in an extensive fashion, except at $\rho_G=0$.  As $\mu$ is increased, the ground state evolves through local rearrangements describable as "micro" first-order transitions\cite{schwab}.  Despite this, there are two qualitatively different regimes where $\delta$ can be either large or small.  In the small $\delta$ regime, the ground state is kinetically accessible, whereas the large $\delta$ regime is characterized by ubiquitous metastability.  If the system becomes stuck in one of these metastable configurations, the geometric frustration will cause anomalously slow evolution towards the ground state.  From Fig.\ref{fig:qvsg2}, it is clear that the symptoms of glassy dynamics become more pronounced once the system reaches quarter filling.

\section{Conclusion}
Having established the main points of the paper it is useful to make a few educated guesses that should be of interest to future work.  We reiterate that as long as the disorder remains finite, it is unlikely that the system will reach a true metallic state.  If the picture described here is generic (ignoring long range Coulomb interaction), there would not be a true metal-insulator transition in $2D$. However, there should be a crossover scale below which the system will appear to have an insulating Mott gap.  In contrast, the corresponding $3D$  case could exhibit a genuine quantum phase transition because the Imry-Ma argument leaves open the possibility of a broken symmetry state signifying a Mott insulator, and the argument for the proliferation of low energy excitations in $D\leq 2$ cannot apply. There are many systems where patchy gapped states or a filled Mott gap appear to be important as in underdoped high temperature superconductors or frustrated magnets. It should be interesting to examine the role of disorder from the present perspective.  Since we have provided a relatively complete characterization of the $1D$ model in the strong interaction limit, it would also be interesting to perform a DMRG simulation where one can explicitly tune the interaction strength.

\acknowledgements
This work is supported by NSF under Grant No. DMR-0705092. We thank Peter Woelfle and S. Shastry for discussions. S. C. would also like to thank the Aspen Center for Physics.

\end{document}